\documentstyle[aps]{revtex}
\begin{document}
\title{Quantum superluminal communication must exist}
\author{Gao Shan}
\address{Institute of Quantum Mechanics, 11-10, NO.10 Building,
YueTan XiJie DongLi, XiCheng District, Beijing 100045, P.R.China.
Tel: 86 010 82882274  Fax: 86 010 82882019  Email:
gaoshan.iqm@263.net}
\date{June 14, 1999}
\maketitle

\vskip 0.5cm

\begin{abstract}

We show that the quantum superluminal communication based on the
quantum nonlocal influence must exist if a basic principle of
science is still valid, which states that if we have demonstrated
the existence of something real, we can confirm its existence.

\end{abstract}

\vskip 0.2cm

After having shown quantum superluminal communication, if exists,
does not result in the causal loop\cite{Gao3}, we will further
demonstrate that quantum superluminal communication must exist due
to the validity of a basic principle of science, which states that
if we have demonstrated the existence of something real, we can
confirm its existence.

At first, as we have demonstrated\cite{Gao3}, the existence of
quantum nonlocal influence requires that there must exist a
preferred Lorentz frame for consistently describing the quantum
nonlocal process, or we will meet logical contradiction in our
explanation about the experimental results about such process,
this conclusion is independent of the existence of the reality
behind the process.

Then, according to a basic principle of science, which states that
if we have demonstrated the existence of something real, we can
confirm its existence, since we have demonstrated the existence of
the preferred Lorentz frame we can find it. This principle is the
only assumption in our demonstrations, and one impressive example
of its validity is the existence of atom.

At last, we will demonstrate that in order to confirm the
existence of the preferred Lorentz frame quantum superluminal
signalling or communication must exist, first, as we have
analyzed\cite{Gao3}, in order to respect special relativity, the
existence of the preferred Lorentz frame, which is only required
by the quantum nonlocal process, will only influence the time
orders of the correlating quantum nonlocal events as in one Bell
experiment, namely in such a preferred Lorentz frame the above
correlating quantum nonlocal events are simultaneous, while in
other inertial frames they are not simultaneous, thus in order to
confirm or find the existence of the preferred Lorentz frame, we
need to identify the time orders of the correlating quantum
nonlocal events in different inertial frames, as to one Bell
experiment this requires that we can distinguish the change of the
measured single state due to the quantum nonlocal influence in
quantum measurement in order to identify the time order of the
change, then this further means that we can distinguish the
nonorthogonal single state, such as $\psi_{1}+\psi_{2}$ and
$\psi_{1}$, which can be easily used to achieve the quantum
superluminal communication, thus we demonstrate that in order to
confirm the existence of the preferred Lorentz frame quantum
superluminal communication must exist.

\end{document}